# Emergent Moiré fringes in direct-grown quasicrystal

Jingwei Li[1,2,#], Kejie Bao[3,4,#], Honglin Sun[1], Xingxu Yan[5,*], Ting Huang[1], Qicheng Zhang[6,7], Yaoqiang Zhou[8], Zhenjing Liu[2], Paul Masih Das[9], Jiawen You[1,2], Jiong Zhao[10], Jianbin Xu[8], Xiaoqing Pan[5], Yongli Mi[2], Junyi Zhu[4,*], Zhaoli Gao[1,*]

1 Department of Biomedical Engineering, The Chinese University of Hong Kong, Hong Kong SAR, China
2 Department of Chemical and Biological Engineering, The Hong Kong University of Science and Technology, Hong Kong SAR, China
3 State Key Laboratory of Surface Physics and Department of Physics, Fudan University, Shanghai 200433, China
4 Department of Physics, The Chinese University of Hong Kong, Hong Kong SAR, China
5 Department of Materials Science and Engineering, University of California - Irvine, Irvine, CA, 92697, USA
6 Research Center for Industries of the Future, Westlake University, Hangzhou, Zhejiang 310030, China.
7. Key Laboratory of 3D Micro/nano Fabrication and Characterization of Zhejiang Province, School of Engineering, Westlake University, Hangzhou, Zhejiang, 310030, China.
8 Department of Electronic Engineering, The Chinese University of Hong Kong, Hong Kong SAR, China
9 Department of Physics and Astronomy, University of Pennsylvania, Philadelphia 19104, USA
10 Department of Applied Physics, The Hong Kong Polytechnic University, Hong Kong SAR, China

#Jingwei Li, Kejie Bao contributed equally to this work
*e-mail: xingxuy@uci.edu, jyzhu@phy.cuhk.edu.hk, zlgao@cuhk.edu.hk

**Abstract**

Quasicrystals represent a category of rarely structured solids that challenge traditional periodicity in crystal materials. Recent advancements in the synthesis of two-dimensional (2D) van der Waals materials have paved the way for exploring the unique physical properties of these systems. Here, we report on the synthesis of 2D quasicrystals featuring 30° alternating twist angles between multiple graphene layers, using chemical vapor deposition (CVD). Strikingly, we observed periodic Moiré patterns in the quasicrystal, a finding that has not been previously reported in traditional alloy-based quasicrystals. The Moiré periodicity, varying with the parity of the constituent layers, aligns with the theoretical predictions that suggest a stress cancellation mechanism in force. The emergence of Moiré fringes is attributed to the spontaneous mismatched lattice constant in the oriented graphene layers, proving the existence of atomic relaxation. This phenomenon, which has been largely understudied in graphene systems with large twist angles, has now been validated through our use of scanning transmission electron microscopy (STEM). Our CVD-grown Moiré quasicrystal provides an ideal platform for exploring the unusual physical properties that arise from Moiré periodicity within quasicrystals.

**Main**

Quasicrystals are a unique class of quasiperiodic ordered materials[1], which challenge the conventional application of Bloch's theorem[2]. Recent studies have shed light on emergent topological properties, as well as optical and electrical behaviors in various quasicrystal systems[3-10], e.g., 1D Fibonacci quasicrystal[11-13], 2D chalcogenide quasicrystal[4,14,15] and 3D icosahedral quasicrystal[16-19]. The assembly of twisted van der Waals materials, such as 30° twisted bilayer graphene (TBLG)[20-23] and tungsten diselenide ($WSe_2$)[24], along with the quasiperiodic systems derived from graphene/hBN/graphene[25,26], has led to the emergence of a new class of 2D quasicrystals with unexplored quantum states. The electronic properties of these materials can be modulated by adjusting the twisted angles and lattice mismatch, giving rise to phenomena such as mirrored Dirac cones[23], quantum anomalous Hall effect[27], and van Hove singularities[24,26]. Recently, the novel concept of "Moiré quasicrystals" achieved by stacking three graphene layers at distinct twist angles[28], twisting bilayer graphene on hBN[29], or creating one–dimensional quasicrystal in the twisted and strained bilayer graphene[30], highlights the significance of establishing an experimental framework that incorporates Moiré periodicity with quasiperiodic order. Such material systems not only facilitate the modulation of electronic interaction within quasicrystals but also promise insights into previously unseen physical phenomena inherent to quasiperiodic systems[25,28-30]. Despite the substantial interest in integrating quasicrystals with periodic Moiré patterns, the direct synthesis of Moiré quasicrystals remains unreported, significantly hindering the studies of the unexplored physics in this unique material system.

In this work, we report a chemical vapor deposition (CVD) process for direct synthesis of Moiré quasicrystal, namely 30° alternating twisted multilayer graphene (30°-ATMG), without the need for additional lattice engineering, and document the presence of Moiré fringes in such CVD-grown graphene quasicrystals. Dark-field transmission electron microscopy (DFTEM) was employed to access the interlayer stacking order in the as-grown 30°-ATMG, where the even layers are twisted by 30° relative to the untwisted odd layers, clearly revealing Moiré fringes in the aligned layer groups. 30° twisted graphene is the quintessential material for studying electronic properties and quantum states in 2D quasicrystals owing to the characteristic relativistic Dirac fermion, which is absent in other TBLG near 30° [20]. We develop a theoretical model, based on the stress cancellation effect, for how the Moiré periodicity in quasicrystalline 30°-ATMG is determined by the parity of graphene layers. This model can be extended to demonstrate the equilibrium conditions for any number of layers, a finding that is corroborated by our experimental observations. The trilayer and pentalayer structures display a Moiré wavelength of ~40 nm, double that of the quadrilayer's ~20 nm. These configurations arise from interlayer energy interactions within the 30°-ATMG, causing compressive or tensile in-plane strains that alter lattice constant and disrupt inversion symmetry. The periodic Moiré fringes, resulting from mismatched lattice constant as validated by scanning transmission electron microscopy (STEM), reveal atomic relaxation in the graphene quasicrystal, a phenomenon previously concealed in studies of 30°-TBLG[20-22]. These discoveries demonstrate that CVD synthesized 30°-ATMG relying on intrinsic mismatched lattice constants provide new avenues for investigating the underexplored physics of quasicrystal.

**Growth and characterization of 30°-ATTLG**

**Figure 1a** shows the optical image of the CVD-grown 30° alternating twisted trilayer graphene (30°-ATTLG) quasicrystal, transferred from a Cu substrate onto a 285 nm thick $SiO_2$/Si substrate. Using the back-diffusion growth mode[31,32], we successfully grew quasicrystalline multilayer graphene on Cu substrates, starting with the largest top layer, followed by a second layer twisted by 30°, and the smallest third layer mirroring the first at the bottom. Our previous report demonstrated that this 30° alternating twisting configuration is energetically favorable[32] and can be identified through optical microscopy (see **Figure S1**).

To explore the stacking order of graphene quasicrystal, Raman spectroscopy was performed on each layer of the 30°-ATTLG (**Figure 1b**). The spectra show that the 2D peak at ~2695 cm$^{-1}$ exhibits a single Lorentzian peak with an intensity surpassing the G peak across all three regions (highlighted by blue, red, black dots in **Figure 1a**), a feature that is especially distinct in the trilayer regions, which aligns with findings from previous studies[23,33]. Additionally, we identify a double-resonance (DR) mode[34,35] in the 30°-ATTLG, with the R peaks in both the bilayer and trilayer regions positioned at ~1379 cm$^{-1}$. The intensity of these R peaks increases with each additional twisted layer, implying the enhanced Umklapp scattering process[23], as evidenced by Raman mapping data (**Figure 1c**).

To further investigate the interlayer atomic registry, the 30°-ATTLG was transferred onto a TEM grid for DFTEM analysis. **Figure 1d** shows the selected area electron diffraction (SAED) pattern from the trilayer region, featuring two sets of six-fold patterns rotated by 30° relative to each other. The diffraction spots corresponding to the 1st/top and 3rd/bottom layers can be clearly distinguished from the 2nd/middle layer due to their significantly higher intensity - nearly twice that of the 2nd layer (**Figure S2**). Moreover, the intensity ratio ($I_1/I_2$) of ~ 0.9 between the first ($Ø_1$) and second order ($Ø_2$) diffraction spots excludes the typical AB or AA stacking configurations between the 1st and 3rd layers, which would exhibit intensity ratios of around 0.3 or greater than 1, respectively[36-38].

The first-order diffraction spot, highlighted by the red circle, was then selected to obtain DFTEM image of the overlap region between the 1st and 3rd layers, strikingly revealing uniform Moiré fringes as shown in **Figure 1e** and **1f**. The presence of these fringes with a periodicity of ~42 nm (**Figure S3**) challenges the traditional view that Moiré fringes do not occur in quasicrystals. To exclude the possibility that these Moiré superlattices result from rotationally faulted graphene layers, several factors were considered: **1)** A periodicity of 42 nm typically correlates with a twisted angle of ~0.34°, which is expected to produce a tessellation of triangular domains with mirrored symmetry (AB/BA)[39]. However, such tessellation is absent in the Moiré fringes observed in our 30°-ATTLG. **2)** The lack of satellite peaks around Bragg peaks in TEM diffraction patterns (**Figure S4**) dismisses the possibility of errors in the twist angle. We, therefore, attribute the formation of the Moiré superlattice in 30°-ATTLG to mismatched lattice constants in the oriented graphene layers, rather than to rotationally faulted graphene layers.

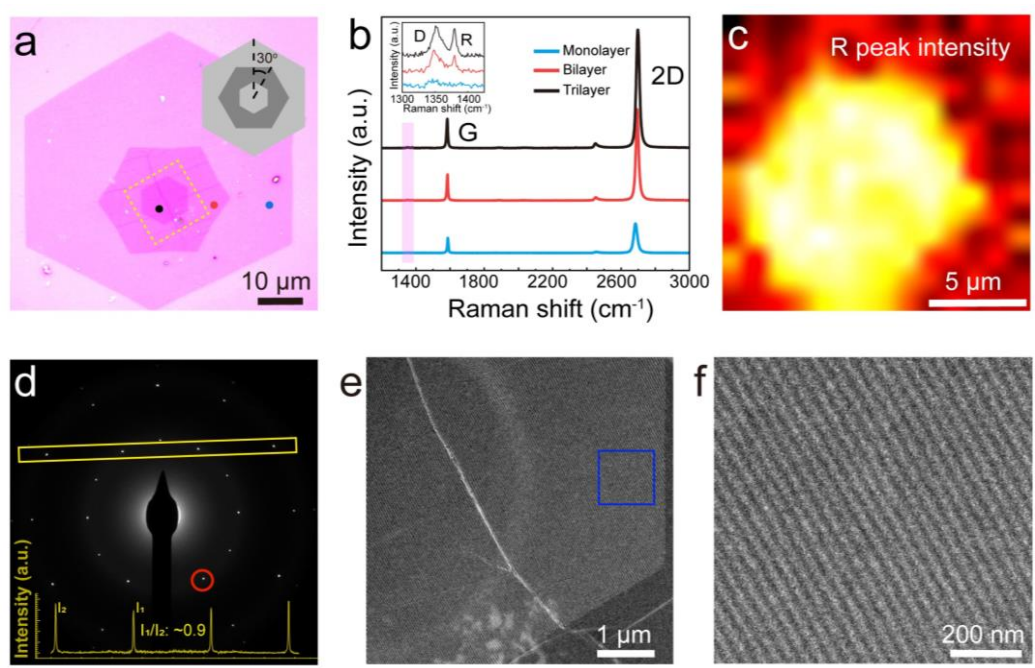

**Figure 1. Characterization of CVD-grown 30°-ATTLG. a** Optical micrograph of 30°-ATTLG transferred onto a 285 nm thick $SiO_2$/Si substrate. The inset illustrates the configuration of 30°-ATTLG, where the 2nd/middle layer is rotated by 30° relative to both the 1st/top and 3rd/bottom layers. **b** Raman spectra from different regions of the sample: monolayer (blue dot), bilayer (red dot), and trilayer region (black dot) in panel **a**. The 2D band at ~2695 cm$^{-1}$ shows a single Lorentzian peak in all three regions, with its intensity exceeding that of the G peak. The inset is the low-energy region (pink-shaded region) of these Raman spectra, highlighting the intervalley DR Raman modes, marked by R peaks at ~1379 cm$^{-1}$ in the bilayer and trilayer areas. **c** Raman mapping of the yellow dashed square area in panel **a** for R peaks, the higher intensity in the trilayer region suggests an enhancement of Umklapp scattering process with the increase of layer number. **d** The SAED pattern of 30°-ATTLG displays two distinct sets

of diffraction peaks, and the set with higher intensity originates from the 1st and 3rd layer. The inset presents the intensity profile for the 1st and 3rd layer set (highlighted by the yellow box), revealing an intensity ratio ($I_1/I_2$) of approximately 0.9 between the first and second order diffraction spots. **e** DFTEM image acquired from the SAED peak (denoted by the red circle in panel **d**). Uniform and periodic Moiré fringes are found in the oriented layers of 30°-ATTLG. **f** A magnified DFTEM image of the highlighted region in panel **e**.

**The Moiré fringes in 30°-ATTLG**

The mismatched lattice constants in the oriented layers of 30°-ATTLG are further elucidated by examining the relationship between the direction of Moiré fringes and their corresponding diffraction peaks[40,41]. This analysis, as depicted in **Figure 2a,** utilized color-composite images from diffraction spots, assigned red, blue, and green, showing that the directions of the Moiré fringes are parallel to their corresponding diffraction peaks. This parallelism indicates an isotropic change in the lattice constants within the oriented layers[40,42]. The wavelength of the Moiré fringes obtained from $Ø_1$ is 42 nm, $\sqrt{3}$ times that of the Moiré fringes obtained from $Ø_2$ (25 nm), as shown in **Figure S5**. This result provides further evidence of a compressed or stretched lattice constant in the 1st and 3rd layers, respectively. To ensure that the observed Moiré superlattice is not influenced by the supporting substrate (carbon film), the sample was transferred onto a TEM grid covered with a holey carbon film. In **Figure 2b**, the Moiré pattern observed in both the hole and out-of-hole regions were nearly identical, with a Moiré wavelength measured at 43 nm (**Figure 2c**), matching the 42 nm observed in the sample supported by the continuous carbon film (**Figure S3**), which dismisses the effect of the substrate. These observations provide evidence that the Moiré fringes, resulting from the mismatched lattice constants in the oriented layers, are an intrinsic feature of the 30°-ATTLG quasicrystal.

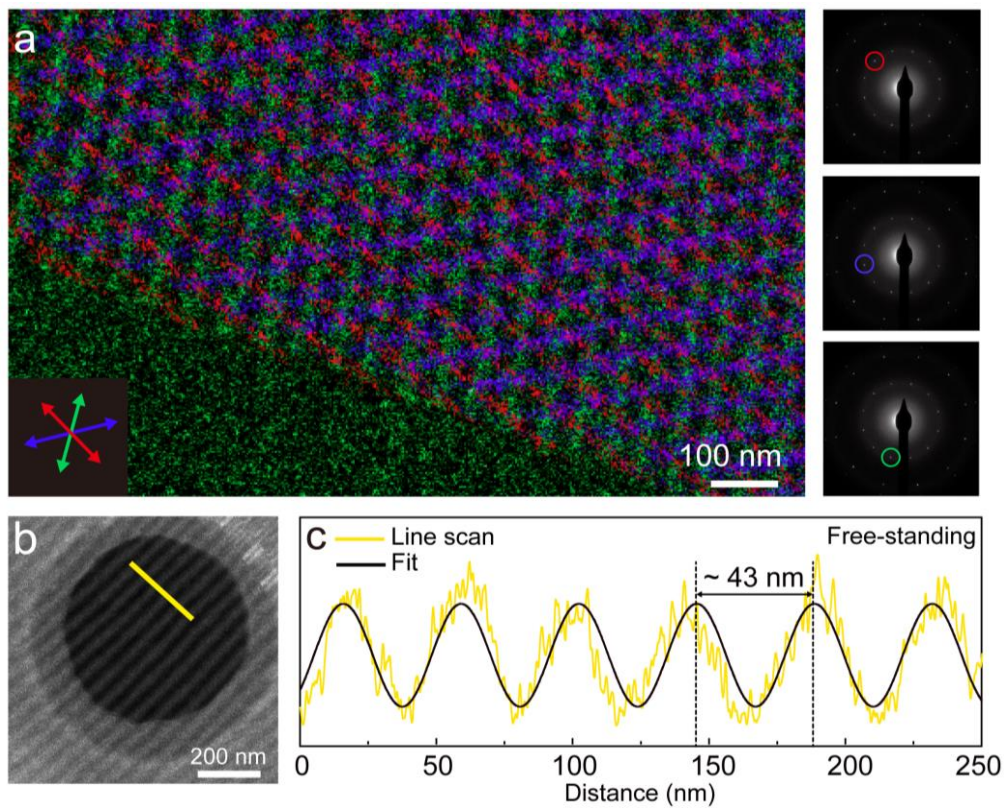

**Figure 2. Visualization of Moiré superlattice in 30°-ATTLG. a** Superimposed DFTEM images of 30°-ATTLG, each obtained from their corresponding diffraction peaks as shown in the right panels, overlaid in red, blue, and green, respectively. The orientation of Moiré fringes aligns with the corresponding diffraction peak, demonstrating an isotropic change in the lattice constant of the oriented layers. **b** DFTEM image shows the periodic Moiré fringes of 30°-ATTLG in the suspended region. The presence of identical Moiré fringe in both the suspended and support regions confirms that substrate effects do not influence the formation of the Moiré pattern. **c** Line profile analysis of the Moiré superlattice in panel **b** reveals a wavelength of approximately 43 nm, matching the 42 nm Moiré periodicity observed in the supported sample (**Figure S3**).

**Stress cancellation effect in 30°-ATMG**

Interlayer coupling can be the main driving force to induce lattice mismatch in 30°-ATTLG Moiré quasicrystal, while such an effect is concealed in a previously studied 30°-TBLG[20-22], as well as 30° non-alternating twisted trilayer graphene (30°-nATTLG). As shown in **Figure S6**, the absence of Moiré fringes in either type of 30°-nATTLG, which exhibit characteristic dislocation patterns, indicates alternating AB/AC stacking changes[42] instead of Moiré patterns. To capture the core mechanism of the Moiré fringes, we constructed a simplified model by treating each layer as a single unit instead of focusing on the individual atomic behaviors, which is effective in weakly coupled two layers. In such a scenario, the equilibrium condition is determined by the minimization of the combination of intralayer strain energy and interlayer binding energy, which may distort the initial lattice. Generally, from the perspective of the layer unit, two stress balance conditions should be satisfied. Firstly, as the whole system is isolated, the internal equilibrium should be satisfied, resulting in the cancellation of the total stress. Secondly, for each middle layer, mediated by the interlayer coupling, the relative stress from two adjacent layers should be balanced if only the nearest layer interaction is considered. Combining the two requirements, the stress distribution of each layer is described by (see **Supplementary Note 2**):

$$\sigma_1 = -(n-1)\delta; \sigma_2 = -(n-3)\delta; \cdots; \sigma_{n-1} = (n-3)\delta; \sigma_n = (n-1)\delta, \quad (1)$$

where $\sigma_i$ is the total stress of each layer, and $\delta$ is a stress unit that is determined by the stacking details. While the trivial solution $\delta = 0$ is also available here, the solution of $\delta \neq 0$ is energetically favored in most cases, such as 30°-ATTLG. To determine the $\delta$ in 30°-ATTLG, we first investigated the behavior of the interlayer binding energy in 30°-TBLG under different strain conditions for simplicity. Because of the lack of translational symmetry, a cluster model with a circular boundary of radius (R) = 100 nm was constructed, as shown in **Figure 3a.** Clearly, the interlayer binding energy for the strain-free lattice is not the global minimum, and multiple local minimums exist for the stained structure. Specifically, the interlayer binding energy is significantly lower for 0.5% compressive or 0.8% tensile strain. If the energy gain from the interlayer part is larger than the energy penalty from the intralayer part, the strain-free lattice could be driven to the strained lattice, leading to $\delta \neq 0$ and lattice mismatch. Such a conclusion can be directly generalized to 30°-ATMG, leading to the lattice mismatch observed in the DFTEM images of 30°-ATTLG.

The uneven distribution of interlayer binding energy against external layer strain indicates the existence of the metastable state with nonzero $\delta_{min}$, as indicated by **Figure 3a.** Interestingly, the $\delta_{min}$ can't be further divided because of the discontinuation of the interlayer energy-strain pattern in 30°-ATMG. We then infer that the solutions of the odd and even numbers for equation (1) are different, if $\delta_{min}$ is considered to be undivided. Specifically, for odd layer number *n*, we have:

$$\sigma_1 = -\frac{n-1}{2}\delta_{min}; \sigma_2 = -\frac{n-3}{2}\delta_{min}; \cdots; \sigma_{n-1} = \frac{n-3}{2}\delta_{min}; \sigma_n = \frac{n-1}{2}\delta_{min}. \quad (2)$$

For the even layer *n*, we have:

$$\sigma_1 = -(n-1)\delta_{min}; \sigma_2 = -(n-3)\delta_{min}; \cdots; \sigma_{n-1} = (n-3)\delta_{min}; \sigma_n = (n-1)\delta_{min}. \quad (3)$$

Therefore, from equation (2) and (3), in 30°-ATTLG, we have:

$$\sigma_1 = -\delta_{min}; \sigma_2 = 0; \sigma_3 = \delta_{min}. \quad (4)$$

The stress distribution in Equation (4) indicates the opposite strain environment for the 1st layer and 3rd layer, which is stretched or compressed in relation to the 2nd layer and thus results in the formation of Moiré fringes.

Moreover, for 30° alternating twisted quadrilayer graphene (30°-ATQLG) and 30° alternating twisted pentalyer graphene (30°-ATPLG), we can obtain:

$$\sigma_1 = -3\delta_{min}; \sigma_2 = -\delta_{min}; \sigma_3 = \delta_{min}; \sigma_4 = 3\delta_{min}. \quad (5)$$

$$\sigma_1 = -2\delta_{min}; \sigma_2 = -\delta_{min}; \sigma_3 = 0; \sigma_4 = \delta_{min}; \sigma_5 = 2\delta_{min}. \quad (6)$$

Obviously, the stress difference between the adjacent layer with the same rotation angle of odd-layer systems is about $2\delta_{min}$, while that of even-layer systems is about $4\delta_{min}$. Quantitively, the wavelength of Moiré fringes can be theoretically predicted. For 30°-ATQLG, both Moiré fringes formed by 1st/3rd and 2nd/4th layer are about 12.3 nm. For 30°-ATPLG, the Moiré fringes formed by the 1st/3rd, 3rd/5th, and 2nd/4th are about 24.6 nm, while those formed by the 1st/5th are about 12.3 nm. The Moiré fringes of the 1st/3rd or 3rd/5th layers and the 1st/5th layers further intertwine, resulting in complex interference fringes (see **Supplementary Note 2**). We proceed to validate the theoretical predictions by further examining the atomic configuration of the oriented layers in both 30°-ATQLG and 30°-ATPLG by DFTEM, as illustrated in **Figure S7** and summarized as **Figure 3b**. The investigation reveals that the 30°-ATQLG comprises two Moiré unit cells and Moiré fringes with a periodicity of approximately 20 nm are observed in both the 1st/3rd and 2nd/4th regions, which is half the wavelength of those observed in 30°-ATTLG. Moiré fringes are also found in 30°-ATPLG, with the Moiré fringes formed by the 2nd/4th layers exhibiting a similar periodicity of around 40 nm to that of 30°-ATTLG. It's worth mentioning that DFTEM image from the 1st/3rd/5th layers exhibit two distinct patterns: one arises from the overlapping region of the 1st/3rd/5th layers, while the other originates from the 1st/3rd layers. Within the Moiré fringes observed in the overlapping 1st/3rd/5th layers region, two types of fringes can be identified, with one appearing relatively brighter and the other darker. These fringes alternate throughout the region, forming a new pattern that results in an increased periodicity of approximately 20 nm. Additionally, Moiré patterns with a periodicity of 40 nm were observed in 1st/3rd layers, resembling the Moiré pattern observed in 30°-ATTLG. The experimentally observed dependence of periodicity on the layer number of the 30°ATMG system aligns with the theoretical predictions.

**Observation of lattice mismatch in Moiré quasicrystal**

To further validate the theoretical model, we employed the low-voltage aberration-corrected annular dark-field (ADF) scanning transmission electron microscopy (STEM) to directly capture the stacking configuration of the 30°-ATTLG at atomic resolution. Based on the relationship between the experimentally observed Moiré periodicity (~40 nm) and its corresponding lattice mismatch (0.6%)[43], as well as the theoretical proposed stress cancellation effect, the 1st and 3rd layers were simulated to exhibit 0.3% tensile and compressive strain, respectively, relative to the 2nd layer for demonstrating the formation of Moiré fringes as shown in **Figure 3c.** Consequently, a transition region of the bridge between AB and AA stacking was determined. In the experimental STEM image (**Figure 3d**), the lattice is observed to be stretched and shifting along a uniaxial direction, compared to the quasicrystalline structure in bilayer graphene with a 30° twist angle[20,21]. The image simulation was further conducted with different amplitude of shift along the [1-210] direction to mimic the transition between AA stacking region to the bridge (**Figure S8**). With the increase of shift from 0 to $0.5a$, where $a$ represents the graphene lattice constant, the atomic structure changes gradually from the regular AA stacking with a classic dodecagonal pattern to bridge with a distorted dodecagonal pattern, and $0.4a$ shift exhibited the most similar stacking feature of the bridge (see **Figure 3e**), while the absence of shift resulted in AA stacking (see **Figure 3f**). In **Figure 3g** and **3h**, the consistent lattice constants observed between two adjacent bright spots in both the experimental (**Figure 3d**) and simulated (**Figure 3e**) STEM images confirm the validity of the aforementioned hypothesis regarding the stretched or compressed lattice constant in relation to the middle layer. The transition stacking in the Moiré fringes was also validated through the intensity ratio of the $\emptyset_1$ and $\emptyset_2$ spots from SAED of 30°-ATTLG. As summarized in **Figure S9**, the $I_1/I_2$ ratios from six 30°-ATTLG samples, ranging between 0.8 and 0.9, differ from those observed in AA or AB stacking and further confirm the stacking transition and distorted lattice constants in the oriented layers.

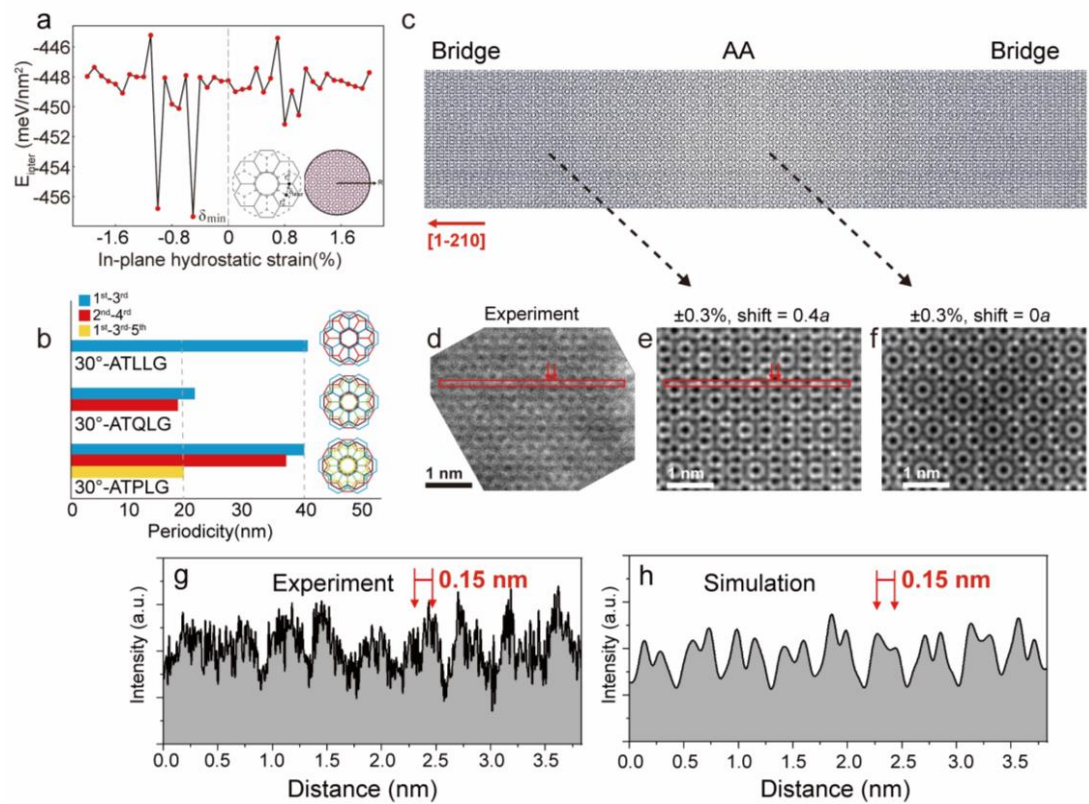

**Figure 3. Atomic relaxation in 30°-ATMG Moiré quasicrystal. a** Relationship between the average intralayer binding energy against the in–plane strain of the top layer. The left inset figure illustrates the interlayer binding energy based on the relative shift of the center of the hexagonal cells in the top and bottom layers. The right inset figure is the schematic illustration of 30°-TBLG cluster with a circular boundary. **b** Summary of the Moiré fringes periodicity observed by DFTEM in 30°-ATMG, highlighting a dependence on the parity of the layer number. **c** Atomic structures of 30°-ATTLG, which demonstrates the transition from bridge region to AA stacking induced by the distorted lattice constant in the 1st and 3rd layer. The lattice of 1st and 3rd layer are stretched and compressed relative to the 2nd layer by 0.3% strain respectively based on the theoretical stress cancellation effect. **d** Atomic–resolution aberration corrected ADF-STEM image taken from the Moiré fringes where the graphene lattice appears stretched. **e, f** Simulated STEM images for the bridge and AA stacking region

respectively. For panel **e**, in addition to the deformed lattice, the 3$^{rd}$ layer has $0.4a$ shift along the [1–210] direction, where $a$ represents the graphene lattice constant. There is no shift of 3$^{rd}$ layer in panel **f**, which exhibits the feature of AA stacking. **g, h** Intensity line profiles obtained from the rectangles in panel **d** (experiment) and **e** (simulation). The atomic distances between two adjacent bright spots are both measured at 0.15 nm; the consistency between experiment and simulation validates the stress cancellation effect.

**Conclusion**

We have demonstrated the direct synthesis of 2D quasicrystals featuring 30° alternating twist angles between multiple graphene layers using CVD, revealing periodic Moiré patterns within the quasicrystal structure - a finding absents in traditional alloy-based quasicrystals. The Moiré periodicity, varying with the parity of the constituent layers, aligns with the theoretical model suggesting a stress cancellation mechanism, where mismatched lattice constants are induced by relative compressive or tensile in-plane strain between adjacent graphene layers. The observed Moiré fringes, attributed to the spontaneous lattice mismatch in oriented graphene layers, were confirmed through STEM, highlighting atomic relaxation within the structure. Our work provides a precise, straightforward, and robust platform for exploring novel physics within Moiré quasicrystals. This unique 2D Moiré quasicrystal, based on van der Waals materials, can be extended to other geometries like dichalcogenides, stimulating the emergence of novel quasicrystal theories and advancing our understanding of quasicrystalline systems.

**Reference:**
1   Lifshitz, R. Symmetry breaking and order in the age of quasicrystals. *Israel Journal of Chemistry* **51**, 1156-1167 (2011).
2   Lesser, O. & Lifshitz, R. Emergence of quasiperiodic Bloch wave functions in quasicrystals. *Physical Review Research* **4**, 013226 (2022).
3   Araújo, R. N. & Andrade, E. C. Conventional superconductivity in quasicrystals. *Physical Review B* **100**, 014510 (2019).
4   Cain, J. D., Azizi, A., Conrad, M., Griffin, S. M. & Zettl, A. Layer-dependent topological phase in a two-dimensional quasicrystal and approximant. *Proceedings of the National Academy of Sciences* **117**, 26135-26140 (2020).
5   Chou, Y.-Z., Wu, F., Sau, J. D. & Sarma, S. D. Correlation-induced triplet pairing superconductivity in graphene-based moiré systems. *Physical Review Letters* **127**, 217001 (2021).
6   Kamiya, K. *et al.* Discovery of superconductivity in quasicrystal. *Nature Communications* **9**, 154 (2018).
7   Man, W., Megens, M., Steinhardt, P. J. & Chaikin, P. M. Experimental measurement of the photonic properties of icosahedral quasicrystals. *Nature* **436**, 993-996 (2005).
8   Vardeny, Z. V., Nahata, A. & Agrawal, A. Optics of photonic quasicrystals. *Nature photonics* **7**, 177-187 (2013).

9   Verbin, M., Zilberberg, O., Lahini, Y., Kraus, Y. E. & Silberberg, Y. Topological pumping over a photonic Fibonacci quasicrystal. *Physical Review B* **91**, 064201 (2015).

10  Weidemann, S., Kremer, M., Longhi, S. & Szameit, A. Topological triple phase transition in non-Hermitian Floquet quasicrystals. *Nature* **601**, 354-359 (2022).

11  Jeon, J. & Lee, S. Topological critical states and anomalous electronic transmittance in one-dimensional quasicrystals. *Physical Review Research* **3**, 013168 (2021).

12  Kraus, Y. E. & Zilberberg, O. Topological equivalence between the Fibonacci quasicrystal and the Harper model. *Physical Review Letters* **109**, 116404 (2012).

13  Moustaj, A., Kempkes, S. & Smith, C. M. Effects of disorder in the Fibonacci quasicrystal. *Physical Review B* **104**, 144201 (2021).

14  Benedek, G., Manson, J. R. & Miret-Artés, S. The electron-phonon interaction of low-dimensional and multi-dimensional materials from He atom scattering. *Advanced Materials* **32**, 2002072 (2020).

15  Park, M. J., Kim, H. S. & Lee, S. Emergent localization in dodecagonal bilayer quasicrystals. *Physical Review B* **99**, 245401 (2019).

16  Hann, C. T., Socolar, J. E. & Steinhardt, P. J. Local growth of icosahedral quasicrystalline tilings. *Physical Review B* **94**, 014113 (2016).

17  Jeon, S.-Y., Kwon, H. & Hur, K. Intrinsic photonic wave localization in a three-dimensional icosahedral quasicrystal. *Nature Physics* **13**, 363-368 (2017).

18  Sinelnik, A. D. *et al.* Experimental observation of intrinsic light localization in photonic icosahedral quasicrystals. *Advanced Optical Materials* **8**, 2001170 (2020).

19  Subramanian, P., Archer, A., Knobloch, E. & Rucklidge, A. M. Three-dimensional icosahedral phase field quasicrystal. *Physical Review Letters* **117**, 075501 (2016).

20  Ahn, S. J. *et al.* Dirac electrons in a dodecagonal graphene quasicrystal. *Science* **361**, 782-786 (2018).

21  Deng, B. *et al.* Interlayer decoupling in 30 twisted bilayer graphene quasicrystal. *ACS Nano* **14**, 1656-1664 (2020).

22  Pezzini, S. *et al.* 30-twisted bilayer graphene quasicrystals from chemical vapor deposition. *Nano Letters* **20**, 3313-3319 (2020).

23  Yao, W. *et al.* Quasicrystalline 30 twisted bilayer graphene as an incommensurate superlattice with strong interlayer coupling. *Proceedings of the National Academy of Sciences* **115**, 6928-6933 (2018).

24  Li, Y. *et al.* Tuning commensurability in twisted van der Waals bilayers. *Nature* **625**, 494-499 (2024).

25  Koshino, M. & Oka, H. Topological invariants in two-dimensional quasicrystals. *Physical Review Research* **4**, 013028 (2022).

26  Leconte, N. & Jung, J. Commensurate and incommensurate double moire interference in graphene encapsulated by hexagonal boron nitride. *2D Materials* **7**, 031005 (2020).

27  Li, Z. & Wang, Z. Quantum anomalous Hall effect in twisted bilayer graphene quasicrystal. *Chinese Physics B* **29**, 107101 (2020).
28   Uri, A. *et al.* Superconductivity and strong interactions in a tunable moiré quasicrystal. *Nature* **620**, 762-767 (2023).
29   Lai, X. *et al.* Imaging Self-aligned Moir\'e crystals and quasicrystals in magic-angle bilayer graphene on hBN heterostructures. *arXiv preprint arXiv:2311.07819* (2023).
30   Sinner, A., Pantaleón, P. A. & Guinea, F. Strain-induced quasi-1D channels in twisted moiré lattices. *Physical Review Letters* **131**, 166402 (2023).
31   Gao, Z. *et al.* Large-area epitaxial growth of curvature-stabilized ABC trilayer graphene. *Nature Communications* **11**, 546 (2020).
32   Gao, Z. *et al.* Crystalline bilayer graphene with preferential stacking from Ni–Cu gradient alloy. *ACS Nano* **12**, 2275-2282 (2018).
33   Kim, K. *et al.* Raman spectroscopy study of rotated double-layer graphene: misorientation-angle dependence of electronic structure. *Physical Review Letters* **108**, 246103 (2012).
34   Carozo, V. *et al.* Raman signature of graphene superlattices. *Nano Letters* **11**, 4527-4534 (2011).
35   Thomsen, C. & Reich, S. Double resonant Raman scattering in graphite. *Physical Review Letters* **85**, 5214 (2000).
36   Lim, H. *et al.* Effects of hydrogen on the stacking orientation of bilayer graphene grown on copper. *Chemistry of Materials* **32**, 10357-10364 (2020).
37   Ping, J. & Fuhrer, M. S. Layer number and stacking sequence imaging of few-layer graphene by transmission electron microscopy. *Nano Letters* **12**, 4635-4641 (2012).
38   Ma, W. *et al.* Interlayer epitaxy of wafer-scale high-quality uniform AB-stacked bilayer graphene films on liquid $Pt_3Si$/solid Pt. *Nature Communications* **10**, 2809 (2019).
39   Yoo, H. *et al.* Atomic and electronic reconstruction at the van der Waals interface in twisted bilayer graphene. *Nature Materials* **18**, 448-453 (2019).
40   Alden, J. S. *et al.* Strain solitons and topological defects in bilayer graphene. *Proceedings of the National Academy of Sciences* **110**, 11256-11260 (2013).
41   Schweizer, P., Dolle, C. & Spiecker, E. In situ manipulation and switching of dislocations in bilayer graphene. *Science Advances* **4**, eaat4712 (2018).
42   Butz, B. *et al.* Dislocations in bilayer graphene. *Nature* **505**, 533-537 (2014).
43   Kim, N. Y. *et al.* Evidence of local commensurate state with lattice match of graphene on hexagonal boron nitride. *Acs Nano* **11**, 7084-7090 (2017).
44   Koch, C. T. *Determination of core structure periodicity and point defect density along dislocations.*  (Arizona State University, 2002).

**Data availability**

The data that support the findings of this study are presented in the Aritcle. Further data are available from the corresponding authors upon reasonable request.

**Acknowledgements**
The work is supported by the Research Grant Council of Hong Kong (Project No. 24201020 and 14207421, AoE/P-701/20) and the National Natural Science Foundation of China (Project No. 62101475). This research is supported in part by project #BME-p2-22 of the Shun Hing Institute of Advanced Engineering, the Chinese University of Hong Kong. The authors acknowledge the use of facilities and instrumentation at the UC Irvine Materials Research Institute (IMRI), which is supported in part by the National Science Foundation through the UC Irvine Materials Research Science and Engineering Center (DMR-2011967).

**Author contributions**
Z.G. and J.Z. directed the project. J.L. conceived and performed the graphene growth experiment. K.B. developed the theoretical model under the supervision of J.Z. TEM characterizaition was performed by P.D, H.S, J.L and Z.L under the supervision of Z.G. ADF-STEM is performed by X.Y. under the supervision of X.P. The TEM results is analyzed by Q.Z., P.D., X.Y., J.Z., J.L and Z.G. TEM samples are preapred by Y.Z., Q.Z., T.H., J.Y. and J.L. under the supervision of J.X.,Y.M. and Z.G. J.L., K.B., X.Y., J.Z., and G.Z. wrote the paper with input and approval from all authors.

**Competing financial interests**
The authors declare no competing financial interests.

**Methods**

**30°-ATMG synthesis and transfer**
The 30°-ATMG was grown by the atmospheric pressure CVD system in 1-in furnace (Lindberg Blue M, Thermo Scientific Co.). Copper (Cu) foils with a thickness of 25 μm (Item #46365, Alfa Aesar) were cleaned by immersing them in 5.4% $HNO_3$ for 1 min, followed by two DI water baths for 1 min each. The Cu foils were then cut into 0.8 cm × 1.5 cm pieces and then loaded into the CVD furnace. The furnace was then heated to a temperature of 1050 °C at a ramping rate of 60 °C/min in a flow of 500 standard cubic centimeters per minute (sccm) of argon (Ar) without hydrogen ($H_2$). This was followed by annealing the growth substrate for 40 min at 1050 °C in a flow of 500 sccm Ar with 30 sccm $H_2$. Multilayer graphene was grown by using 5.2 sccm of 5000 parts per million (ppm) $CH_4$ (diluted with Ar) with 30 sccm $H_2$ for 2 h. The reactor was then rapidly cooled to room temperature in a flow of 10 sccm $H_2$ and 1000 sccm Ar.

The "bubbling" method was utilized to peel the graphene from the Cu substrate. PMMA (MicroChem Corp., PMMA950, A4) was spin-coated onto the Cu foil with deposited graphene and then baked at 105 °C for 2 min. The PMMA-coated Cu foil was gradually immersed into the aqueous solution of NaOH with a concentration of 50 mM and a voltage of 15 V was applied. As hydrogen bubbles formed at the copper surface, a graphene /PMMA film could be peeled off from the Cu foil. The graphene/PMMA film was then rinsed with deionized (DI) water three times and transferred onto a 285 nm $SiO_2$/Si substrate. The chip was allowed to dry naturally and then baked at 150 °C for 2 min before removing PMMA with acetone. The multilayer graphene on $SiO_2$/Si substrate was subsequently annealed in $H_2$/Ar forming gas at 300 °C for 1 h.

**TEM sample preparation**
Optical microscopy was used for identifying target sample transferred on the $SiO_2$/Si substrate, after which gold markers were transferred near the target multilayer graphene sample for precise localization. Subsequently, the adhesion between the gold marker and the substrate was enhanced by 1 h annealing under $H_2$/Ar atmosphere at 300 °C. The graphene on $SiO_2$ was then spin-coated with PMMA once again and etched with a 1M KOH solution at 90 °C for a duration of 1 h. The PMMA/graphene was subsequently cleaned with two water baths, picked up by the TEM grid, and dried for 3 h. Finally, the PMMA on the TEM grid was eliminated by acetone and IPA.

**DFTEM**
The target multilayer graphene with gold markers were transferred on the TEM Cu grid. TEM characterization was performed by FEI Tecnai 12 TEM (120 kV). Under TEM, the gold maker enables a fast locating of the target graphene. An aperture was positioned in the diffraction plane of an electron microscope to generate a real space image from electrons scattered only at a specified angle. The DFTEM technique is highly responsive to the atomic registry of the oriented layers.

**ADF-STEM**
Before putting the as-prepared TEM grids into electron microscopes, the TEM grids were baking in vacuum at 160 °C over 8 h to further remove residual polymers and moisture. Low voltageAtomic resolution aberration corrected ADF STEM imaging was performed using Nion UltraSTEM 200 operating at a 60 kV accelerating voltage. The convergence semi-angle is 38 mrad with a beam current of ~10 pA. The inner and outer collection semi-angles of the ADF detector are 70 and 210 mrad, respectively. STEM image simulations were performed using a multislice-based simulation software called QSTEM[44]. All input parameters, including probe size, convergence angle and collection angles of the ADF detector, were set according to experimental conditions.

**Numerical simulation for 30°-ATMG**
The interlayer binding energy of 30°-ATMG was simulated by a cluster model with circular boundary condition. The interlayer energy between two hexagonal cells of two layers was obtained by the interpolation method between the AA and AB stacking. The interlayer binding energy between the two layers can be treated as the summation of interlayer energy inside the circular cluster. R=100 nm was used in the calculations and different compressive or tensile strains were applied to illustrate the behavior of the interlayer binding energy against external strain. For multilayers, a quantitative model focusing the layer relaxation instead of atomic relaxation was constructed to explain the Moiré fringes in the experiments.